\documentclass[twocolumn,aps,prl,superscriptaddress,tightenlines]{revtex4-2}
\usepackage{amsmath}
\usepackage{amsfonts}
\usepackage{graphicx}
\usepackage{epsfig}
\usepackage{color}
\usepackage{textcomp}
\usepackage{physics}
\usepackage{wasysym}
\usepackage{float}
\usepackage{blindtext}
\usepackage{mathtools}
\usepackage{booktabs}
\usepackage{pifont}
\usepackage{orcidlink}
\definecolor{myblue}{rgb}{0.18039,0.1882353,0.57255}
\definecolor{myred}{rgb}{1,0.,0.3}

\usepackage{mathrsfs}
\usepackage{changes} 
\usepackage{hyperref}
\definechangesauthor[name={Per cusse}, color=red]{per} 
\hypersetup{
	colorlinks=true,
	citecolor=myblue,
	linkcolor=myblue,
	urlcolor=myblue,
}
\begin{document}
	
	\title{Disorder-Induced Exponential Scaling of Subradiant Decay Rates}
	\author{Guoqing Tian\orcid{0009-0000-6801-1361}}
	\affiliation{School of Physics and Institute for Quantum Science and Engineering, Huazhong University of Science and Technology, and Wuhan institute of quantum technology, Wuhan, 430074, P. R. China}
	\author{Xin-You L\"{u}\orcid{0000-0003-3561-6684}}\email{xinyoulu@hust.edu.cn}
	\affiliation{School of Physics and Institute for Quantum Science and Engineering, Huazhong University of Science and Technology, and Wuhan institute of quantum technology, Wuhan, 430074, P. R. China}
	
	\date{\today}
	
	\begin{abstract}
		Subradiance, a hallmark cooperative phenomenon in waveguide QED, is characterized by a universal power-law scaling of decay rates with system size and underpins many applications in quantum information storage. Here, we demonstrate that disorder drives a sharp transition in the typical subradiant decay rates from power-law to exponential scaling, a phenomenon we term the subradiant scaling transition (SST). Through rigorous finite-size scaling analysis, we establish the SST as a critical phenomenon, characterized by a diverging characteristic scale of the decay rates at the transition point $W_c=0$. Physically, the SST originates from Anderson localization, manifested by the physical equivalence between the characteristic scale and the localization length of the subradiant states. Our findings provide deep insights into the interplay between disorder and collective dynamics, unifying the underlying physical mechanisms of exponentially-scaled subradiant decay rates and Anderson localization in waveguide QED.
	\end{abstract}
	\maketitle

	
	In waveguide QED, collective interactions between quantum emitters and their photonic environment give rise to a rich variety of cooperative phenomena\,\cite{corzo_waveguide-coupled_2019,mirhosseini_cavity_2019,reitz_cooperative_2022,sheremet_waveguide_2023}. Notable among these is subradiance\,\cite{sub_exp1,zhang_subradiant_2020,brehm2021waveguide,poshakinskiy_dimerization_2021}, where destructive interference suppresses radiation to produce long-lived excitations, a counterpart to the enhanced emission of superradiance\,\cite{sup_1,sup_2,sup_3,sup_4, masson2022universality}. While quantum emitters coupled to the free-space electromagnetic vacuum exhibit diverse scalings for subradiant decay rates\,\cite{asenjo-garcia_exponential_2017,kornovan_extremely_2019,Zhang2020}, those coupled to a one-dimensional (1D) waveguide share a defining universal feature: their subradiant decay rates follow an $N^{-3}$ law\,\cite{albrecht_subradiant_2019,zhang_theory_2019}. This profound power-law scaling not only unveils the intricate interplay between coherence and dissipation in waveguide QED systems but also underpins various applications in quantum information processing and memory\,\cite{app3,app1,asenjo-garcia_exponential_2017,app2}.

	
	In experimental platforms, disorder arising from fabrication imperfections is inevitable. Such disorder significantly influences transport properties, most notably through Anderson localization, a phenomenon where interference in disordered media halts wave propagation and confines excitations to localized regions\,\cite{anderson_absence_1958, AL2, abrahams_scaling_1979, evers_anderson_2008}. Over the past decade, the interplay between disorder and cooperativity in waveguide QED has emerged as an active research frontier, with growing efforts to understand how disorder influences collective dynamics\,\cite{akkermans_photon_2008, viggiano_cooperative_2023, gjonbalaj_modifying_2024,zhang2025robustsuperradiancespontaneousspin,giant_atom}. While it is known that disorder can generally modify collective emission, its precise impact on subradiance has yet to be fully elucidated. In particular, how the universal $N^{-3}$ scaling of subradiance in waveguide QED responds to disorder remains to be thoroughly investigated. Furthermore, whether a fundamental connection exists between subradiance and the inevitable emergence of Anderson localization\,\cite{JHH, mbl, dis1} is still entirely unexplored. Addressing these questions is not only essential for understanding collective quantum optical phenomena but also crucial for designing practical quantum photonic devices.

	\begin{figure}
		\centering
		\includegraphics[width=8cm]{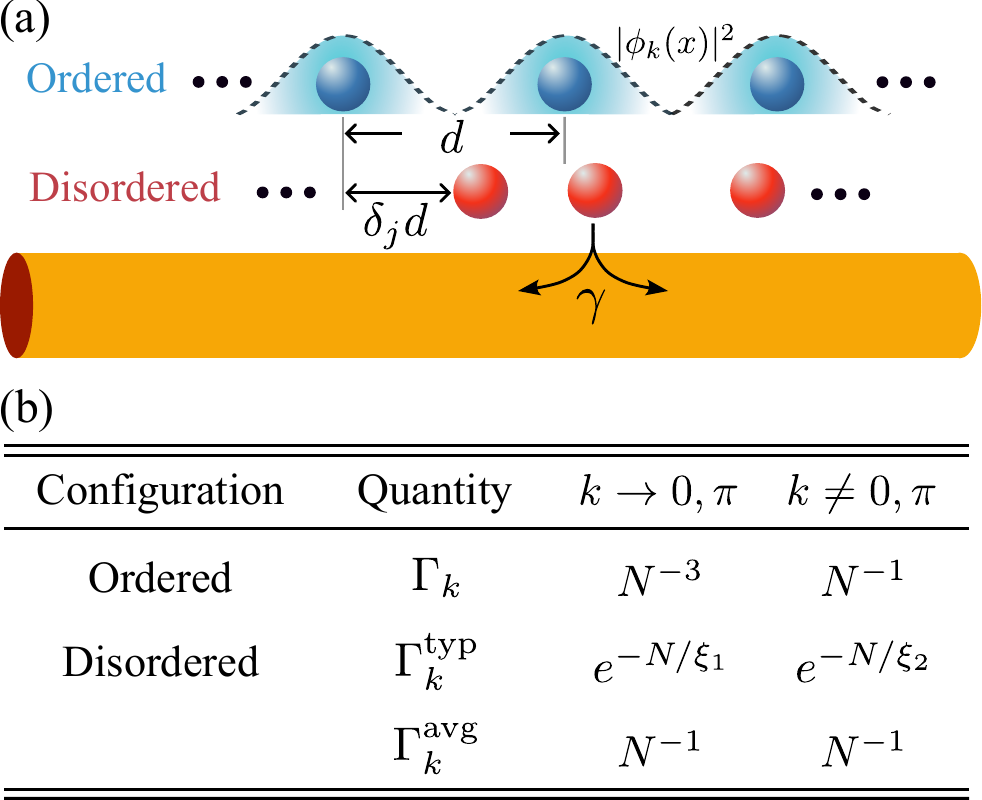}
		\caption{(a) Schematic of a waveguide-coupled qubit array with positional disorder $\delta_j d$. Ordered subradiant states (dashed line) are delocalized standing waves $\phi_k(x)\propto \sin(kx)$. (b) Scaling laws for decay rates in ordered and disordered chains. We compare strong subradiant states (with $k \to 0,\pi$) and weak subradiant states (with $k \neq 0,\pi$).}\label{model}
\end{figure}

\begin{figure}
	\centering
	\includegraphics[width=8.5cm]{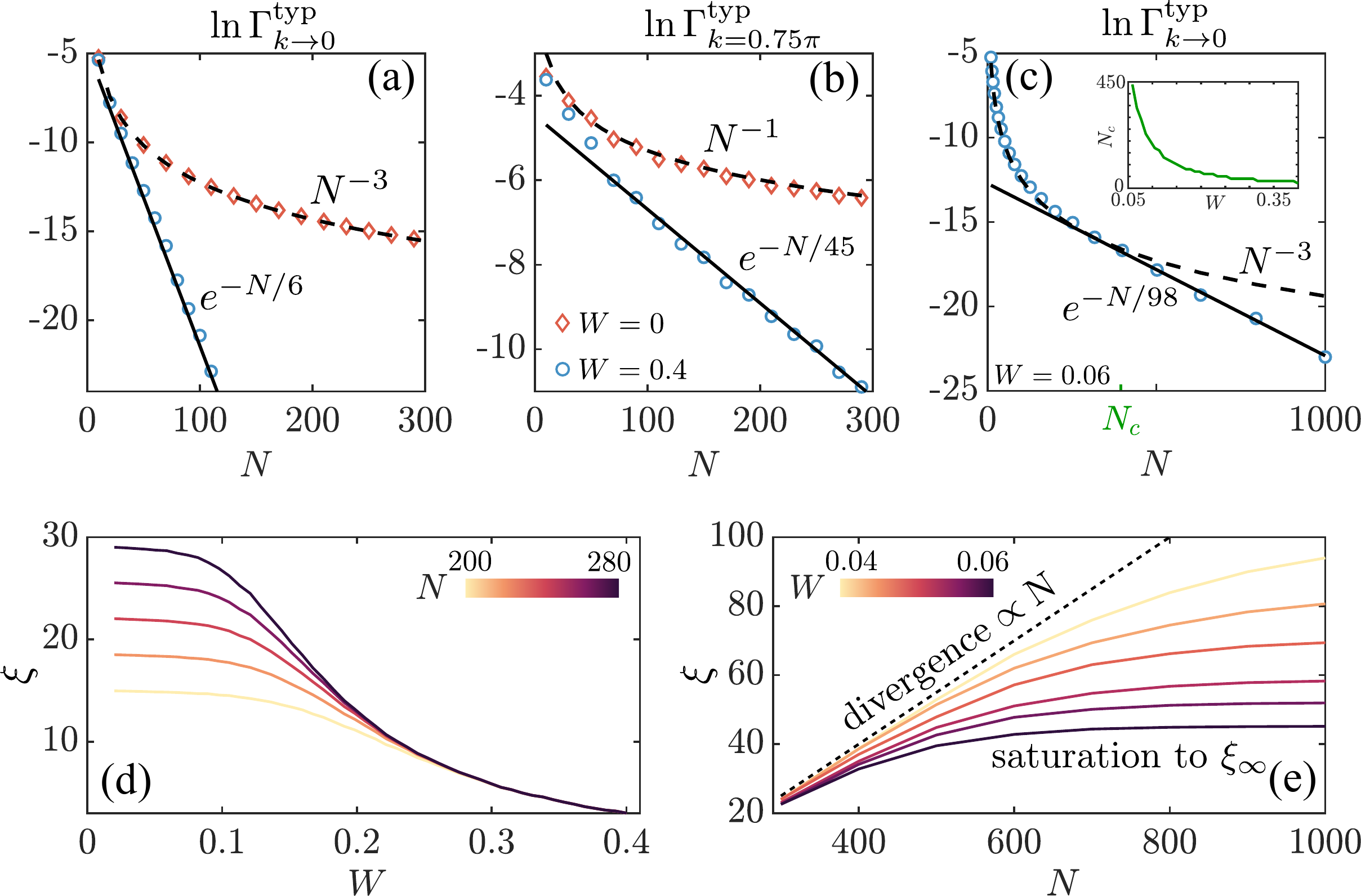}
	\caption{Scaling of $\Gamma^{\rm typ}_{k=0.75\pi}$ [(a)] and $\Gamma^{\rm typ}_{k\to0}$ [(b)] with system size $N$ for $W=0$ (orange diamond) and $W=0.4$ (blue circle). (c) Scaling of $\Gamma^{\rm typ}_{k\to 0}$ at $W=0.06$. Inset shows the dependence of $N_c$ on $W$. In (a-c), solid (dashed) line represents exponential (power-law) fit. (d) Finite-size characteristic scale $\xi$ for $\Gamma^{\rm typ}_{k\to 0}$ versus $W$ for different $N$. (e) The dependence of $\xi$ for $\Gamma^{\rm typ}_{k\to 0}$ on $N$ for different disorder strengths. The black dashed line represents a linear divergence $\propto N$. In (a-e), all results assume $\varphi=0.5\pi$ and are ensemble-averaged over $10^3\sim 10^4$ disorder realizations.}\label{fig2}
	
\end{figure}


We address these questions by investigating the scaling behaviors of two distinct classes of subradiant decay rates $\Gamma_{k}$ in a disordered 1D waveguide QED system [see Fig.\,\ref{model}(a)], which are distinguished by their different $N$-scaling behaviors in ordered systems. Because disorder induces broadly skewed distributions of decay rates, the standard arithmetic mean, $\Gamma^{\text{avg}}_k \equiv \langle \Gamma_k \rangle$, merely reflects the rare events of the disordered system (detailed discussions are provided in\,\cite{supp}). To accurately capture the most probable events, one needs to evaluate the typical decay rate, defined as $\Gamma^{\text{typ}}_k \equiv \exp(\langle \ln \Gamma_k \rangle)$, where $\langle \bullet \rangle$ denotes the disorder average. We analyze both $\Gamma^{\text{typ}}_k$ and $\Gamma^{\text{avg}}_k$, with key findings summarized in Fig.\,\ref{model}(b).

Specifically, for the typical values, disorder induces a scaling transition from algebraic to exponential for both subradiant classes. In contrast, the mean values exhibit divergent behaviors: both the strong and weak subradiant decay rates exhibit $N^{-1}$ scaling. Through rigorous finite-size scaling analysis, we demonstrate that the algebraic-to-exponential scaling transition of the typical values, termed the subradiant scaling transition (SST), exhibits critical nature. This is manifested by a diverging characteristic scale $\xi \propto (W-W_c)^{-\nu}$ near the transition point $W_c=0$, with the critical exponent $\nu$ depending on the specific wavevector $k$. Finally, by examining the spatial profiles of these states, we reveal that the SST originates intrinsically from Anderson localization. We establish this connection by demonstrating a physical equivalence between the exponential scaling of decay rates and the exponential spatial localization of the wavefunctions.

\emph{Model}---We consider a chain of $N$ qubits coupled to a 1D waveguide, as illustrated in Fig.\,\ref{model}(a). In the Markovian limit, the collective dynamics of the qubits are described by the master equation
\begin{align}
	\!\!\!\!\!\dot{\hat{\rho}}=-i[\hat{H}_{\rm eff}\hat{\rho}\!-\!\hat{\rho} \hat{H}^{\dagger}_{\rm eff}]\!+\!\sum_{mn}\gamma\cos(|x_m\!-\!x_n|k_0)\hat{\sigma}_m\hat{\rho}\hat{\sigma}^{\dagger}_n.
\end{align}
Here $\hat{\sigma}_m=|g_m\rangle\langle e_m|$ is the lowering operator for the $m$th qubit at position $x_m$, with $|g_m\rangle$ ($|e_m\rangle$) denoting the ground (excited) state; $\gamma$ represents the decay rate of each qubit, and $k_0=\omega_0/v_g$ is the resonant wavevector. In a rotated frame with respect to $\hat{H}_0=\sum_{m}\omega_0\hat{\sigma}^{\dagger}_m\hat{\sigma}_m$, the non-Hermitian effective Hamiltonian is
\begin{equation}\label{eq1}
	\hat{H}_{\rm eff}=-\frac{i\gamma}{2}\sum_{m,n=1}^{N}e^{i|x_m-x_n|k_0}\hat{\sigma}^{\dagger}_m\hat{\sigma}_n.
\end{equation}
Interference effects render the system highly sensitive to the spatial arrangement of the qubits. We introduce positional disorder by defining the qubit positions as $x_m = md + \delta_m d$, where $d$ is the uniform spacing of the ordered configuration. The dimensionless offsets $\delta_m$ are drawn from a uniform distribution $\delta_m \in [-W/2, W/2]$. The disorder strength $W$ is restricted to $0 \leq W < 1$ to preserve the spatial ordering $x_m > x_n$ (for $m > n$). Given the system symmetry, we restrict our analysis of the parameter $\varphi=k_0d$ to the range $0\le\varphi\le \pi/2$. Hereafter, we set $\gamma=d=1$.


\emph{Algebraic-to-exponential scaling transition}---In the single-excitation subspace, the effective Hamiltonian satisfies the eigenvalue equation $\hat{H}_{\text{eff}}|\phi_k\rangle = \omega_k |\phi_k\rangle$. In the absence of disorder, the eigenvalues are given by $\omega_k = (\gamma/4)[\cot((\varphi+k)/2) + \cot((\varphi-k)/2)]$\,\cite{zhang_theory_2019}, and the eigenstates $|\phi_k\rangle$ approximate Bloch standing waves, with spatial probability distribution $|\phi_k(x)|^2 \sim \sin^2(k x)$. Due to the dissipative nature of the excitations, the quasimomentum $k$ is complex, satisfying $\Re k \in (0, \pi)$ and $\lim_{N\to\infty} \Im k = 0$. The decay rates, $\Gamma_k=-\Im\omega_k$, exhibit distinct $N$-scaling behavior depending on $k$. Specifically, when $k$ approaches the center or the boundary of the Brillouin zone, i.e., $k \to 0$ or $\pi$, the corresponding subradiant decay rates scale as $N^{-3}$\,\cite{albrecht_subradiant_2019,zhang_theory_2019,kornovan_extremely_2019}; for $k \neq 0, \pi$, the decay rates scale as $N^{-1}$\,\cite{supp}. We refer to these two types of subradiant states as strong and weak subradiant, respectively.

	Figures\,\ref{fig2}(a,b) present the scaling of both strong and weak subradiant states with $N$ for ordered and disordered configurations. In contrast to their power-law scaling in ordered arrays, both classes exhibit a disorder-induced transformation. Specifically, for small $N$, the decay rates retain their respective algebraic behaviors ($N^{-3}$ for strong, $N^{-1}$ for weak subradiance). As $N$ increases, however, the scaling in both cases crosses over to an exponential form $\Gamma^{\rm typ}_k \propto \exp(-N/\xi_{\infty})$, characterized by distinct values of $\xi_{\infty}$. Given the qualitatively similar behavior between the strong and weak subradiant decay rates in the presence of disorder, for simplicity, subsequent panels (c)-(e) focus only on the strong subradiant decay rate. This crossover $N$-scaling becomes more pronounced in the weak disorder regime, as shown in Fig.\,\ref{fig2}(c). To quantify this scaling crossover, we define $N_c$ as the system size where $\Gamma^{\rm typ}_{k\to 0}$ deviates from its ordered algebraic counterpart by a factor of $e^{-1}$, while its ratio to the exponential fit exceeds the threshold of $2e^{-1}$. The inset illustrates the dependence of $N_c$ on the disorder strength $W$, showing that $N_c$ increases as disorder decreases.
	
	To robustly extract the characteristic scale $\xi_{\infty}$ of the exponential decay from finite-size data, we first compute the finite-size characteristic scale via $\xi=\frac{M_3}{M_2}-\frac{M_2}{M_1}$, where $M_q=\sum_n^N n^q\Gamma^{\rm typ}_k(n)$ is defined as the discrete $q$-rd moment, and $\Gamma^{\rm typ}_k(n_i)$ is the decay rate for a chain of size $n_i$. The thermodynamic-limit characteristic scale is then defined as $\xi_{\infty}=\lim_{N\to\infty}\xi$. Figure\,\ref{fig2}(d) displays $\xi$ as a function of disorder strength $W$ for various system sizes. In the strong disorder regime, $\xi$ remains largely independent of $N$. Conversely, for weak disorder, $\xi$ exhibits a pronounced size dependence, increasing monotonically with $N$. This behavior, further detailed in Fig.\,\ref{fig2}(e), stems directly from the crossover scaling of $\Gamma^{\rm typ}_{k\to 0}$. In the regime $N \ll N_c$, the dominance of power-law scaling manifests as a linear divergence of $\xi$ with $N$. As the system size exceeds the crossover scale ($N \gg N_c$), the emergence of exponential scaling causes $\xi$ to saturate to its constant thermodynamic-limit value $\xi_{\infty}$.
	
	\begin{figure}
		\centering
		\includegraphics[width=8.5cm]{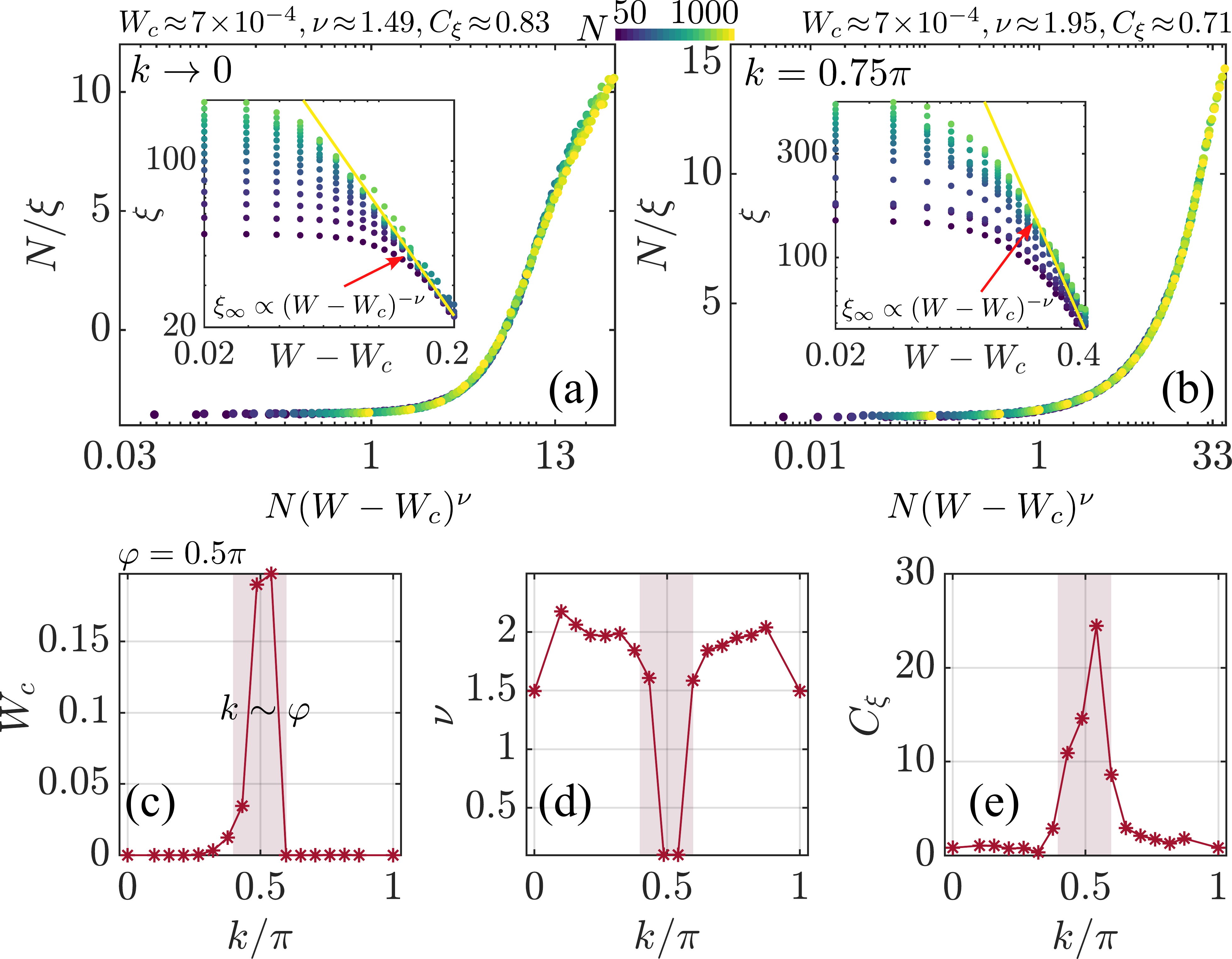}
		\caption{Data collapse of the finite-size characteristic scale extracted from $\Gamma^{\rm typ}_{k\to 0}$ [$\Gamma^{\rm typ}_{k=0.75\pi}$] in (a) [(b)]. The insets show the dependence of $\xi$ on $(W-W_c)$ for different $N$. The yellow solid lines represent the fit to $\xi_{\infty}\propto (W-W_c)^{-\nu}$ using the critical parameters obtained from the collapse. Critical disorder strength [(c)], critical exponent [(d)], and cost function [(e)] versus $k$. The shaded area schematically marks the vicinity of $k=\varphi$. In (a-e), $\varphi=0.5\pi$. All results are ensemble-averaged over $10^3$ disorder realizations. The details about data collapse are provided in\,\cite{supp}.}\label{fig3}
	\end{figure}
	
	\emph{Criticality analysis}---The size-dependent behavior of $\xi$ is a clear signature of finite-size effects, strongly suggesting an underlying criticality of $\xi_{\infty}$. Beyond these finite-size manifestations, the criticality of $\xi_{\infty}$ is further evidenced by the fundamentally distinct $N$-scaling behaviors in the ordered versus disordered limits. In the ordered chain ($W=0$), the pure power-law scaling of $\Gamma^{\rm typ}_k$ indicates $N_c\to \infty$, and thus $\xi_{\infty}\to \infty$. In contrast, in the presence of sufficient disorder, $\Gamma^{\rm typ}_{k}$ follows an exponential scaling at $N\gg N_c$, resulting in the finite $\xi_{\infty}$. This disparity implies the existence of a critical disorder strength $W_c$ marking the transition where $\xi_{\infty}$ shifts from diverging to finite. This corresponds to a fundamental transition in the scaling of decay rates from algebraic to exponential in the thermodynamic limit, a phenomenon we term the SST.
	
	We assume that the critical behavior of $\xi_{\infty}$ near $W_c$ is given by $\xi_{\infty} \propto (W - W_c)^{-\nu}$, where $\nu$ denotes the critical exponent. To accurately extract these critical parameters from numerical data of finite systems, we perform a finite-size scaling (FSS) analysis\,\cite{fss1,fss2}. According to FSS theory, the critical properties of $\xi_{\infty}$ near the critical point are also reflected in the finite-size characteristic scale $\xi$. Specifically, if $\xi_{\infty}$ follows the critical behavior described above, $\xi$ for different $N$ and $W$ should take the universal scaling form $N/\xi = \mathcal{F}\left[ N(W-W_c)^\nu \right]$, where $\mathcal{F}(\bullet)$ is a universal scaling function. This universal behavior across different parameters can be further verified by a data collapse analysis. If $\xi$ follows this universal scaling form, plotting $N/\xi$ against $N(W - W_c)^\nu$ for all data points will cause the curves for different $N$ to collapse onto a single master curve. The quality of the collapse is quantified by the cost function $C_{\xi}$ (defined in\,\cite{FSS_supp1,supp}), where a value closer to zero indicates a better collapse. 
	
	Figures\,\ref{fig3}(a) and (b) present the data collapse for representative strong ($k\to 0$) and weak ($k=0.75\pi$) subradiant states, respectively, demonstrating that the $N/\xi$ data fall well ($C_{\xi}\lesssim 1$) onto corresponding master curve. The extracted critical disorder strength are $W_c\approx 0$ in both cases, while $\nu\approx 1.49$ ($\nu\approx 1.95$) for the strong (weak) subradiant state. The insets further confirm the validity of these collapses. As $N$ increases (finite-size effects vanish), $\xi$ exhibits the predicted divergence $(W-W_c)^{-\nu}$, implying $\xi_{\infty}$ also approaches the same divergence. Figures\,\ref{fig3}(c,d) further show the extracted critical parameters for various $k$. For strong subradiant states, the critical exponent is approximately given by $\nu \approx 1.5$. Conversely, for weak subradiant states, $\nu$ increases rapidly to approximately $2$. Physically, the smaller critical exponent associated with strong subradiance implies that, near the critical point, disorder suppresses the radiative decay of strong subradiant states more effectively than that of their weak counterparts. Remarkably, we find $W_c \approx 0$ for both strong and weak subradiant decay rates, indicating that the power-law scaling of subradiant decay rates in the ordered chain is fundamentally fragile: the introduction of extremely small disorder immediately destroys the power-law scaling of all subradiant decay rates, triggering the transition to exponential. Notably, when $k$ is near $\varphi$, $C_{\xi}$ becomes significantly large [$C_{\xi}\gg 1$, see Fig.\,\ref{fig3}(c)], indicating that our criticality assumption is invalid in this regime. This is because the states corresponding to $k \sim \varphi$ are in fact superradiant, which do not undergo the SST and thus exhibit no critical behavior\,\cite{supp}.

\begin{figure}
	\centering
	\includegraphics[width=8.5cm]{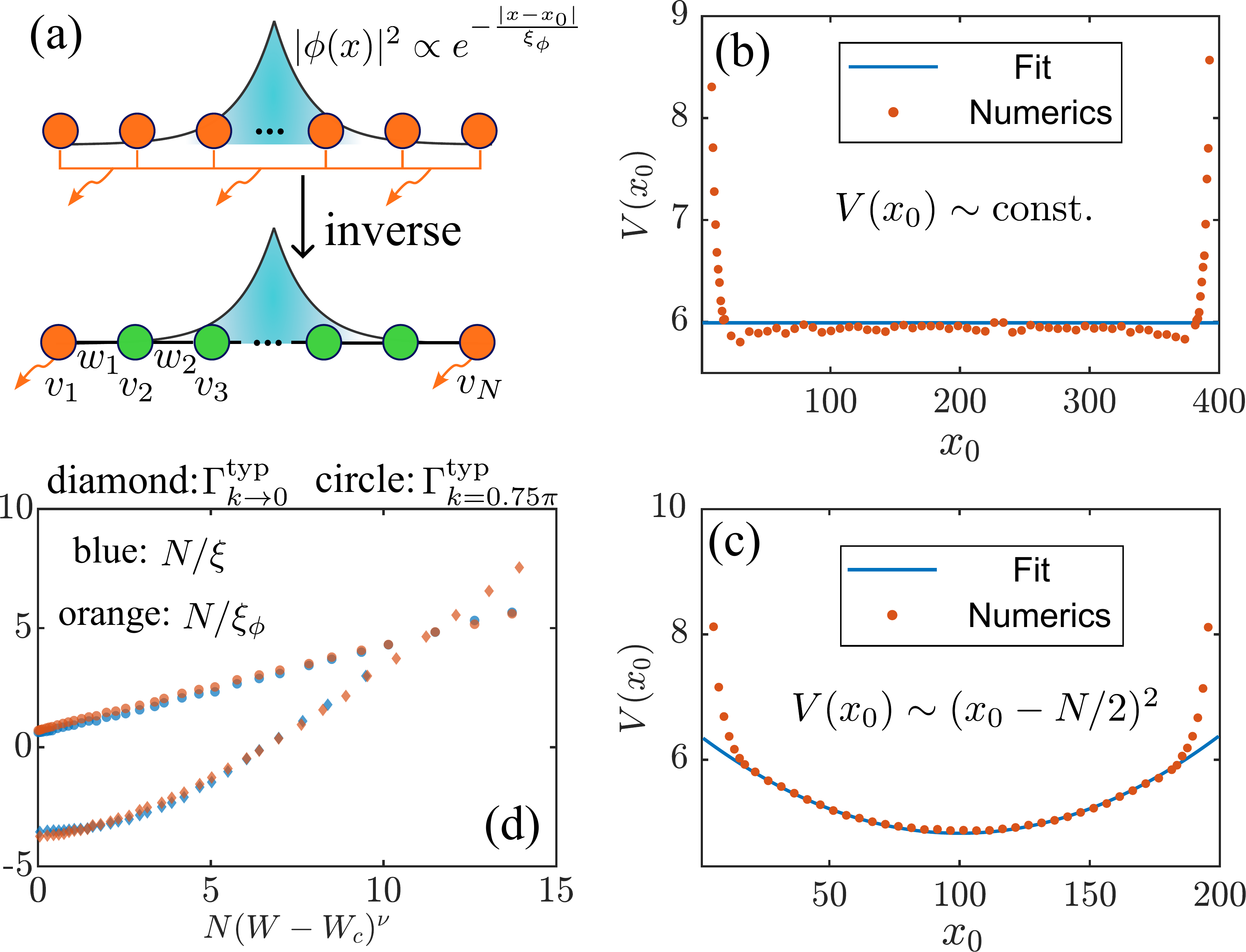}
	\caption{(a) Schematics of the Hamiltonian configuration in the single-excitation subspace for $\hat{H}_{\rm eff}$ (top) and $\hat{H}^{-1}_{\rm eff}$ (bottom). Disordered subradiant states (solid line) are localized wavepacket $\phi(x)\propto e^{-|x-x_0|/\xi_{\phi}}$. (b-c) Effective potential ($N=400$) obtained from the strong subradiant state in (b), and for the weak subradiant state (with $k=0.75\pi$ and $N=200$) in (c). $W=0.2$. The blue lines denote the numerical fits. (d) Data collapse of the localization length for subradiant state with $k\to 0$ (blue diamond) and $k=0.75\pi$ (blue circle). Data collapse of the characteristic scale for $\Gamma^{\rm typ}_{k\to 0}$ (orange diamond) and $\Gamma^{\rm typ}_{k=0.75\pi}$ (orange circle). The extracted critical exponents are $\nu_{\phi}=1.51$ for strong subradiant state and $\nu_{\phi}=1.93$ for weak subradiant state. The critical disorder strength are $W^{\phi}_c\approx 0$ for both cases. In (b-d) all results assume $\varphi/\pi=0.5$.}\label{fig4}
\end{figure}

\emph{The connection between SST and Anderson localization}---The critical properties of the SST, particularly the critical point at $W_c \approx 0$, bear a striking resemblance to those of Anderson localization in low-dimensional systems. This strong similarity implies a possible connection between the underlying mechanism of SST and Anderson localization. To explore this connection, we first examine the inverse of the effective Hamiltonian, $\hat{H}_{\text{eff}}^{-1}$\,\cite{Inverse_1,sheremet_waveguide_2023,Inverse_2}. Within the single-excitation subspace, it is given by $\hat{H}^{-1}_{\rm eff}=\hat{H}_0+i\hat{V}$. Here, $\hat{H}_0=\sum_{m=1}^{N}v_m|m\rangle\langle m|+\sum_{m=1}^{N-1}w_m(|m\rangle\langle m+1|+|m+1\rangle\langle m|)$ and $\hat{V}=\gamma^{-1}(|1\rangle\langle 1|+|N\rangle\langle N|)$, where $v_1/\gamma=-\cot(\Delta \varphi_1/2)$, $v_N/\gamma=-\cot(\Delta \varphi_{N-1}/2)$, $v_{1<m<N}/\gamma=-\cot(\Delta \varphi_m)-\cot(\Delta \varphi_{m-1})$, and $w_m/\gamma=-\csc(\Delta \varphi_m)$ with $\Delta \varphi_m=k_0(x_{m+1}-x_m)$. One advantage of working with the inverse Hamiltonian is that it transforms the physical picture of collective emission into boundary radiation, as schematically shown in Fig.\,\ref{fig4}(a). This can be seen by noting the spectral decomposition $\hat{H}_{\text{eff}}^{-1}\ket{\phi_k} = \omega_k^{-1}\ket{\phi_k}$, which leads to ${\rm Im}\langle \phi_k|\hat{H}^{-1}_{\rm eff}|\phi_k\rangle=\Gamma_k/(\Gamma^2_k+\Omega^2_k)=2(|\phi_k(1)|^2 + |\phi_k(N)|^2)/\gamma$, with $\Omega_k=\Re\omega_k$. For a subradiant state with $\Gamma_k \ll |\Omega_k|$, this relation simplifies to $\Gamma_k \approx \tilde{\gamma} (|\phi_k(1)|^2 + |\phi_k(N)|^2)$, where we define $\tilde{\gamma}=2\Omega^2_k/\gamma$. Therefore, the decay rate of a subradiant state $|\phi_k\rangle$ is essentially determined by its population at the boundaries. Furthermore, in this inverted picture, the originally infinite-range dipole-dipole interactions are mapped onto a tight-binding model with only nearest-neighbor hopping. This mapping greatly simplifies the analysis of the spatial distribution of eigenstates.

Within the inverse picture, introducing positional disorder transforms $\{v_m\}$ and $\{w_m\}$ into random variables, while the boundary dissipation mechanism and the tight-binding structure remain unaltered. A key feature of this system is that these parameters exhibit strictly finite-range spatial correlations\,\cite{supp}. For 1D Hermitian systems, it is well-established that such short-range correlated disorder drives a transition from extended to localized states\,\cite{AL3,AL4}, also known as the Anderson localization. For the non-Hermitian Hamiltonian $\hat{H}^{-1}_{\rm eff}$, we show that Anderson localization still occurs\,\cite{supp}. This is because, although the effective inverse Hamiltonian $\hat{H}^{-1}_{\rm eff}$ is non-Hermitian, the restriction of non-Hermiticity to the boundaries acts as a perturbation that does not alter the bulk localization properties significantly. This indicates that the Bloch waves of the ordered system collapse into spatially localized modes.

In the presence of disorder that induces localization, we model the spatial distribution as a wavepacket localized at $x_0$ with localization length $\xi_\phi$, i.e., $\phi(x) \propto \exp(-|x - x_0|/\xi_\phi)$. The boundary population is then given by $(|\phi(1)|^2+|\phi(N)|^2)\propto \exp(-N/\xi_{\phi})\cosh((2x_0-N)/\xi_{\phi})$. Based on the boundary radiation mechanism, the typical decay rate of this localized state is given by
\begin{equation}\label{bd}
	\Gamma^{\rm typ} \propto \exp[\int_{1}^{N}\ln(e^{-\frac{N}{\xi_{\phi}}}\cosh(\frac{2x_0-N}{\xi_{\phi}}))P(x_0)\dd{x_0}].
\end{equation}
Here, the stochastic behavior of the decay rates stems entirely from the randomness of the wave-packet center $x_0$. To characterize this randomness, we introduce the probability density function $P(x_0)$ for $x_0$ and define an effective potential via $V(x_0) = -\ln P(x_0)$. Figures\,\ref{fig4}(b) and (c) show $V(x_0)$ for the strong and weak subradiant states, respectively. For the strong subradiant state, we find $V(x_0)$ is well described by a constant value. As for the weak subradiant states, $V(x_0)$ is well fitted by a harmonic-like soft potential $(x_0 - N/2)^2/\sigma(N)^2$, where $\sigma(N)$ denotes the effective width. Crucially, we find that this width expands superlinearly with the system size, $\sigma(N) \propto N^\alpha$ ($\alpha > 1$)\,\cite{supp}. This superlinear growth ensures that the confining potential becomes asymptotically flat in the large-$N$ limit, thus statistically reducing to a constant potential akin to that of the strong subradiant states. Note that, in both cases, $V(x_0)$ drops rapidly near the edges and thus deviate significantly from the fitted values. This boundary depletion arises because the open boundary conditions truncate the exponential tails of the localized states, making such edge-adjacent configurations statistically unfavorable. Based on the fitted distributions $P(x_0)$, we analytically demonstrate that both classes of subradiant decay rates universally follow the exponential scaling $\Gamma \propto \exp(-N/2\xi_{\phi})$\,\cite{supp}. This explicitly identifies the extracted characteristic scale as $\xi = 2\xi_{\phi}$, thereby establishing a profound unification: the spectral characteristic scale $\xi$ governing the radiative lifetime is intrinsically equivalent to the spatial localization length $\xi_{\phi}$, i.e., $\xi\equiv\xi_{\phi}$.

The equivalence between the characteristic scale and the localization length physically stems from the combined effect of the boundary dissipation mechanism and Anderson localization. Since the wavepacket centers are statistically uniformly distributed across the bulk, a typical subradiant state localizes at a distance proportional to the system size $N$ from the boundaries. Consequently, this typical localized state must tunnel through a macroscopic distance to radiate at the edges, naturally leading to an exponential suppression. This equivalence also explains why the critical disorder strength for the SST is found at $W_c = 0$, a property inherited directly from Anderson localization in low dimensions, where any infinitesimal disorder is sufficient to suppress extended transport. To further confirm this equivalence, we perform a data collapse analysis in Fig.\,\ref{fig4}(d), comparing the localization length $\xi_{\phi}$ with the characteristic scale $\xi$. The localization length is extracted from the participation ratio via $\xi_{\phi}=\left(\sum_x |\phi(x)|^4\right)^{-1}$. The obtained critical parameters are very close to those of the characteristic scale $\xi$. Moreover, the data for both quantities, after a conformal transformation that preserves the critical behavior, collapse well onto a single master curve, providing strong numerical evidence for their intrinsic equivalence.


\emph{Conclusion}---In summary, we have unveiled a disorder-induced subradiant scaling transition in waveguide QED, where the typical decay rates transition from an algebraic to an exponential scaling with system size. We demonstrated that this transition stems from Anderson localization, physically governed by the tunneling of center-pinned localized wavepackets to the dissipative boundaries. Beyond the typical behavior, we revealed that rare boundary events lead to a power-law scaling laws for the means. This work bridges the gap between cooperative radiative phenomena and disordered localization physics, offering a unified framework for understanding subradiance in realistic quantum systems.

\emph{Acknowledgement}---We appreciate Tao Shi and Qing-Yang Qiu for their insightful suggestions regarding the calculations about the decay rates and the finite-size scaling analysis. We also thank Hua-Jin Gao, Liang-Liang Wan, and Zhi-Guang Lu for their valuable suggestions to our work. This work was supported by the National Science Fund for Distinguished Young Scholars of China (Grant No. 12425502) and the National Key Research and Development Program of China (Grant No. 2021YFA1400700). The computation was completed on the HPC Platform of Huazhong University of Science and Technology.

\bibliography{test}
\end{document}